\documentstyle[multicol,aps,prl,twocolumn]{revtex}
\input{psfig}
\begin{document}
\draft 
\topmargin=0.01cm
\twocolumn[\hsize\textwidth\columnwidth\hsize\csname@twocolumnfalse\endcsname

\title{Intermittency in passive scalar advection}
\author{U.~Frisch$^{1}$, A.~Mazzino$^{1,2}$ and M.~Vergassola$^{1}$\\
\small{$^{1}$ CNRS, Observatoire de la C\^ote d'Azur, B.P. 4229,
06304 Nice Cedex 4, France.}\\
\small$^2$ INFM--Dipartimento di Fisica, Universit\`a di Genova, I--16146
Genova, Italy.}
\date{\today}
\maketitle
\begin{abstract}
A Lagrangian method for the numerical simulation of the Kraichnan
passive scalar model is introduced. The method is based on
Monte--Carlo simulations of tracer trajectories, supplemented by a
point-splitting procedure for coinciding points.  Clean scaling
behavior for scalar structure functions is observed.
The scheme is exploited to investigate the
dependence of scalar anomalies on the scaling exponent $\xi$ of the
advecting velocity field. The three-dimensional fourth-order
structure function is specifically considered.

\end{abstract}
\pacs{PACS number(s)\,: 47.10.+g, 47.27.-i, 05.40.+j}]

The Kraichnan model of passive scalar advection by a self-similar Gaussian
white-in-time velocity field\, \cite{RHK94} is by now a paradigm for
intermittency in turbulent systems.  It was shown by
R.H.~Kraichnan\, \cite{RHK68} that, for advecting velocity fields
which are white-in-time ($\delta$-correlated), the equal-time
correlation functions of the scalar field $\theta$ obey closed equations
of motion. In his 1994 paper\, \cite{RHK94} he used this remarkable property
and the so-called ``linear ansatz''  to predict the scaling exponents
$\zeta_n$ of the
$n$-th order scalar structure function $S_n$ for all space dimensions
$d\ge2$ and for all velocity scaling exponents $0\le\xi\le 2$. The 
predicted $\zeta_n$'s are {\em
anomalous}, for example $2\zeta_2-\zeta_4>0$; in other words the scaling
exponents cannot be predicted by dimensional analysis.  Hence, the
flatness of scalar increments over a distance $r$  grows $\propto
r^{\zeta_4-2\zeta_2}$ as $r\to 0$, a phenomenon referred to as
``intermittency'' \cite{UF95}. The linear  ansatz was revisited via fusion
rules in Ref.~\cite{FGLP96} and its validity in the limit $\xi\to 0$ was
questioned in Ref.~\cite{RHK97}. A different approach was developed in
Refs.~\cite{CFKL95,GK95,SS95} in which anomalous scaling has its roots in
the  zero modes (solutions of the unforced problem) of the closed exact
equations satisfied by the scalar correlations \cite{nonlinear}.
Their determination for correlation functions with more than two or
three point presents a daunting task which has so far received
solutions only via perturbation theory. Three limits have been identified
for the Kraichnan model\,: large $d$'s
\cite{CFKL95}, small
$\xi$'s
\cite{GK95} and $\xi$'s close to the Batchelor limit $\xi=2$
\cite{SS95,SS96}. The first two expansions are regular, while for the
third one the relevant small parameter should be $\sqrt{2-\xi}$. (This
is due to the preservation of the collinear geometry in the Batchelor
limit, leading to an angular non-uniformity in the perturbation
analysis \cite{SS95,SS96}.)

Numerical simulations have up to now been based on the direct
integration of the passive scalar partial differential equation and
have been limited to two dimensions\,\cite{KYC95,FGLP97}. Although the
predictions of the linear ansatz appear compatible with such
simulations, it should be noted that such calculations are highly
delicate. To wit, the difficulty of observing for $S_2$ the known
asymptotic scaling \cite{RHK94}.

Our aim here is to propose a different numerical strategy based on the
Monte--Carlo simulation of Lagrangian trajectories\,\cite{GPZ97}. For
structure functions of finite order only a finite number of tracer
particles is needed. The tracer trajectories are easy to simulate
and the calculation at each time step only involves a small number of
random vectors, basically, differences of velocities, rather than the whole
velocity field. Furthermore, working with the tracers naturally allows to
measure the scaling of the structure functions
$S_{2n}^{(L)}(r)=
\langle
\left(\theta({\mbox{\boldmath $r$}})-\theta(0)\right)^{2n}\rangle$ 
{\it vs} the integral scale $L$ of the forcing. Physically, this means
that the passive scalar variance injection rate (which equals its
dissipation rate) and the separation $r$ are kept fixed while the
integral scale $L$ is varied.  In an anomaly-free theory, e.g. of the
Kolmogorov 1941 type, nothing should change in inertial-range
statistical quantities. Anomalies will here be measured directly
through the scaling dependence on $L$ of the structure functions.

Specifically, let us consider the passive scalar equation
\begin{equation}
\label{passive}
\partial_t\theta({\mbox{\boldmath $r$}},t)+{\mbox{\boldmath $v$}}
({\mbox{\boldmath $r$}},t)\cdot{\mbox{\boldmath $\nabla$}}\,
\theta({\mbox{\boldmath $r$}},t)=\kappa\nabla^2
\theta({\mbox{\boldmath $r$}},t)+f({\mbox{\boldmath $r$}},t) .
\end{equation}
For the Kraichnan model \cite{RHK94}, the velocity and the forcing are
Gaussian independent processes, both homogeneous, stationary, isotropic
and white-in-time. The velocity is self-similar  (no
infrared cutoff is needed nor assumed in our procedure); the
correlations of its increments are given by\,:
\begin{eqnarray}
\label{correlations}
\langle v_{\alpha}({\mbox{\boldmath $r$}},t)
\,v_{\beta}({\mbox{\boldmath $r$}},0)
\rangle -\langle v_{\alpha}({\mbox{\boldmath $r$}},t)\,v_{\beta}(0,0) \rangle =
\nonumber \\
r^{\xi}\left[\left(\xi+d-1\right)\delta_{\alpha\beta}-\xi\,{r_{\alpha}
r_{\beta}\over r^2}\right]\delta(t)\, .
\end{eqnarray}
As for the forcing, $\langle f({\mbox{\boldmath $r$}},t)\,f(0,0)\rangle =
F\left(r/L\right)\,\delta(t)$, where $F\left(r/L\right)$ 
is nearly  constant for
distances $r$ smaller than the integral scale $L$ and decreases rapidly for
$r\gg L$.

When the molecular diffusivity $\kappa$ is simply ignored, and
$\theta$ is assumed to vanish in the distant past at $t=-T$,
eq.~(\ref{passive}) can be recast as $\theta({\mbox{\boldmath $r$}},t)
=\int^t_{-T} f({\mbox{\boldmath $r$}}(s),s)\,ds$, with the Lagrangian
trajectory defined by the stochastic differential equation 
$d{\mbox{\boldmath $r$}}(s)={\mbox{\boldmath $v$}}
\left({\mbox{\boldmath $r$}}(s),s\right)\,ds$ and the final condition
${\mbox{\boldmath $r$}}(t)={\mbox{\boldmath $r$}}$.  
Using the Wick rule to calculate Gaussian
averages over the forcing, the scalar correlations can be expressed as
averages of time integrals of $F$ over the statistics of Lagrangian
trajectories. Furthermore, zero-mode ideas suggest the universality of
the scaling exponents with respect to the choice of $F$.  It is then
convenient to consider the step function $F=1$ for $r\leq L$ and $F=0$
for $r > L$. (The fact that its Fourier transform is not positive
definite is not relevant for the sequel as it actually amounts to
taking a complex forcing function.) The scalar correlations have then
very simple expressions, e.g. for the second and the fourth-order
correlations in the stationary state\,:
\begin{eqnarray}
\label{times}
 &\langle \theta({\mbox{\boldmath $r$}}_1)\,\theta({\mbox{\boldmath $r$}}_2)
\rangle= \langle T^{L}_{12}
\rangle_{\cal L}\,; &   \\
&\langle \theta({\mbox{\boldmath $r$}}_1)\,\theta({\mbox{\boldmath $r$}}_2)
\theta({\mbox{\boldmath $r$}}_3)\theta({\mbox{\boldmath $r$}}_4)
\rangle = \langle T^{L}_{12}T^{L}_{34}+T^{L}_{13}T^{L}_{24}
+T^{L}_{14}T^{L}_{23} \nonumber &
\rangle_{\cal L}.
\end{eqnarray}
Here, $T^{L}_{12}$ is the (random) amount of time that two particles
starting at ${\mbox{\boldmath $r$}}_1$ and ${\mbox{\boldmath $r$}}_2$
and moving backwards in time spend with their mutual distance
$|{\mbox{\boldmath $r$}}_1(s)-{\mbox{\boldmath $r$}}_2(s)| < L$ and
$\langle\bullet\rangle_{\cal L}$ denotes the average over the
Lagrangian trajectory statistics. Expressions similar to (\ref{times})
are easily derived for higher order correlations.  Note that we can
exchange backward and forward motion in time since, according to
(\ref{correlations}), the statistical properties of ${\mbox{\boldmath
$v$}}$ and $-{\mbox{\boldmath $v$}}$ are the same.

In the limit $\kappa\to 0$ this procedure, which ignores molecular
diffusion, is actually correct as long as all points 
${\mbox{\boldmath $r$}}_i$ are
distinct. However, if we, e.g.,  put ${\mbox{\boldmath $r$}}_1
={\mbox{\boldmath $r$}}_2$ we find that
$\langle \theta^2\rangle$, given by (\ref{times}) is incorrect\,: it
diverges
$\propto T$ as $T\to \infty$. With coinciding points, the correct procedure
is the point-splitting\,: the tracer particles must be initially separated
by a small distance
$O(\epsilon)$ and the value of the correlation function for coinciding
points is given by the limit
$\epsilon\to 0$. This is finite for any $\xi < 2$ because, even for
$\epsilon\to 0$, the particles reach an $O(1)$ separation  in a finite
time, on account of the H\"older non-smooth nature of the velocity field
(see, e.g., Ref.~\cite{BGK97} for this important property of what may be
termed a ``Richardson walk''). It is then  easily checked that
$\langle\theta^2\rangle$  coincides with the known value at
$\kappa=0$ of the analytical solution \cite{RHK94} and that for
$\xi=2$ its divergence is logarithmic in $\epsilon$, as it should be in
the Batchelor regime.

In our simulations, the  point-splitting operation is most conveniently
implemented by keeping a non-vanishing amount of ``molecular noise''.
By this we understand that the different
particles, in addition to being swept along by the velocity field, are
undergoing {\em independent\/} Brownian motions with a small diffusivity
$\kappa$.  This Brownian diffusion is relevant only for interparticle
distances smaller than the dissipation scale
$\eta=O\left(\kappa^{1/\xi}\right)$. The corresponding  stochastic
equations of motion for the case of  $2n$  tracer particles are\,:
\begin{eqnarray}
\label{noisy}
d{\mbox{\boldmath $r$}}_i(s)&=& {\mbox{\boldmath
$v$}}\left({\mbox{\boldmath $r$}}_i(s),s\right)\,ds+\sqrt{2\kappa}\,\,
{\dot {\mbox{\boldmath $W$}}}_i(s)\,ds\,,\;\; i=1,\ldots ,2n,\nonumber\\ 
{\mbox{\boldmath $r$}}_i(t)&=&{\mbox{\boldmath $r$}}_i
\end{eqnarray}
where $\langle \left({\dot W}_i\right)_{\alpha}(s)\,\left({\dot
W}_j\right)_{\beta}(s')\rangle=
\delta_{ij}\delta_{\alpha\beta}\delta(s-s')$ and ${\mbox{\boldmath
$r$}}_i$ is the position at the (final) time $t$.  It can be checked
that the $2n$-th order scalar correlation functions given by
(\ref{times}), with ${\mbox{\boldmath $r$}}_i$ and $t$ interpreted as
Eulerian coordinates, satisfy Kraichnan's closed equations (for
details, see Refs.~\cite{BGK97,MC97}).

The Lagrangian method defined by (\ref{times}) and (\ref{noisy}) is
numerically implemented as follows. Because of homogeneity only
differences in positions and velocities matter and we can work with
$2n-1$ particles for the moments of order $2n$.  The $2n$-th order
structure function $S_{2n}$ requires $n+1$ configurations of such
particles. For example, $S_2(r)=2\left[\langle\theta^2\rangle-
\langle\theta(r)\theta(0)\rangle\right]$. Since the velocity field is
white-in-time, equations such as (\ref{noisy}) could present the well
known Ito--Stratonovich ambiguity\, \cite{KP92} which is however
absent as a consequence of incompressibility.  The tracer positions
are updated using the classical Euler--Ito scheme of order $1/2$
\cite{KP92}. Thus, during the time interval $\Delta s$ the Lagrangian
positions for each configuration of tracers $\left(r_i\right)_\alpha$
($i=1,\ldots , 2n-1$ and $\alpha=1,\ldots , d$) are shifted by
$\sqrt{\Delta
s}\left(\left(X_i\right)_\alpha+\left(Y_i\right)_\alpha\right)$. Here,
$\left(X_i\right)_{\alpha}$ and $\left(Y_i\right)_{\alpha}$ are two
sets of $\left(2n-1\right)d$ Gaussian random variables chosen
independently at each time step and with the appropriate correlations.
For example, using ${\mbox{\boldmath $r$}}_2-{\mbox{\boldmath
$r$}}_1,\ldots ,{\mbox{\boldmath $r$}}_{2n}-{\mbox{\boldmath $r$}}_1$
as dynamical variables, we have $\langle\left(X_1\right)_1
\left(X_2\right)_3\rangle = \langle \left(v'({\mbox{\boldmath $r$}}_2)
-v'({\mbox{\boldmath $r$}}_1)\right)_1 \left(v'({\mbox{\boldmath
$r$}}_3)-v'({\mbox{\boldmath $r$}}_1)\right)_3\rangle$ and
$\langle\left(Y_1\right)_1 \left(Y_2\right)_1\rangle=2\kappa$. (The
$v'$-field has no time dependence and the same spatial correlations as
the $v$-field.) Individual realizations are stopped when all the
interparticle distances become larger than ten times the largest
integral scale of interest $L_{max}$. The number of realizations
needed for the results reported below is from one to several millions.

A severe test for the Lagrangian method is provided by the
second-order structure function $S_2$, whose expression is known
analytically \cite{RHK94}.
Its behavior being non-anomalous, a flat
scaling in $L$ should be observed.
\begin{figure}
\begin{center}
\vspace{-0.7cm}
\mbox{\hspace{0.0cm}\psfig{file=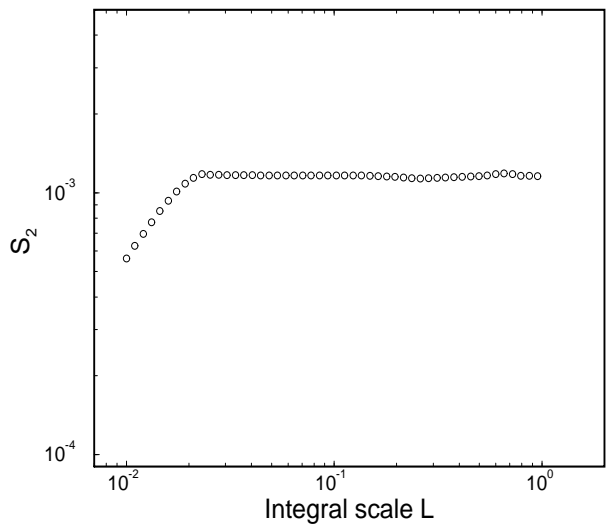,height=8cm,width=9cm}}
\end{center}
\vspace{-0.6cm}
\caption{The 3-D second-order structure function
$S_2$ {\it vs} $L$ for $\xi=0.6$. The separation $r=2.7\times 10^{-2}$,
the diffusivity $\kappa=1.115\times 10^{-2}$ and the number of realizations
is $4.5\times 10^6$.}
\label{fig1}
\end{figure}
The structure function $S_2$
measured by the Lagrangian method is shown in Fig.~\ref{fig1} for
$\xi=0.6$ and $d=3$ (all structure functions are plotted in log--log
coordinates).
The measured slope is $10^{-3}$ and the error on the
constant is $3\%$. (These figures are typical also for other values of
$\xi$ studied.)
Two remarks are in order. First, it follows from the
analytic solution, that the constant-in-$L$ behavior holds for
all $r < L$, including in the dissipative region. Physically, this
corresponds to the fact that, as $r$ moves down in the dissipative region,
the energy flux becomes smaller and smaller, but  still remains
independent of $L$. Second, the asymptotic scaling for  $L\ll r$ goes
over into the scaling for $\langle\theta^2\rangle$, namely
$L^{2-\xi}$;  the transition to the constant-in-$L$ behavior around $r=L$
is very sharp, on account of the step function chosen for $F$.
\begin{figure}
\begin{center}
\mbox{\hspace{0.0cm}\psfig{file=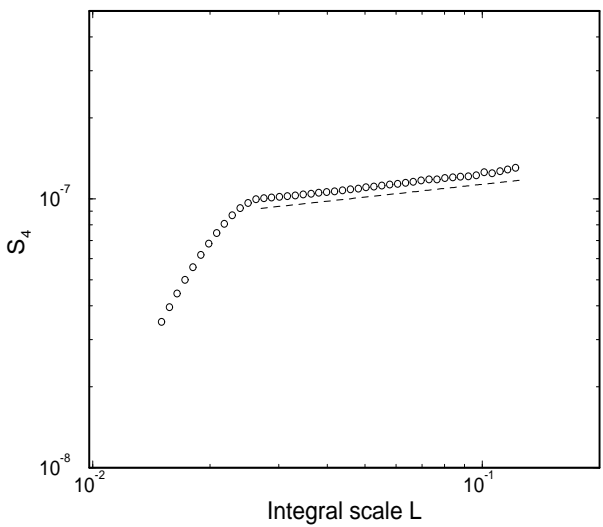,height=8cm,width=9cm}}
\end{center}
\vspace{-0.5cm}
\caption{The 3-D fourth-order structure function $S_4$ {\it vs} $L$ for
$\xi=0.2$. The separation $r=2.7\times 10^{-2}$, the
diffusivity $\kappa=0.247$ and
the number of realizations is $15\times 10^6$.}
\label{fig3}
\end{figure}

\begin{figure}
\begin{center}
\mbox{\hspace{0.0cm}\psfig{file=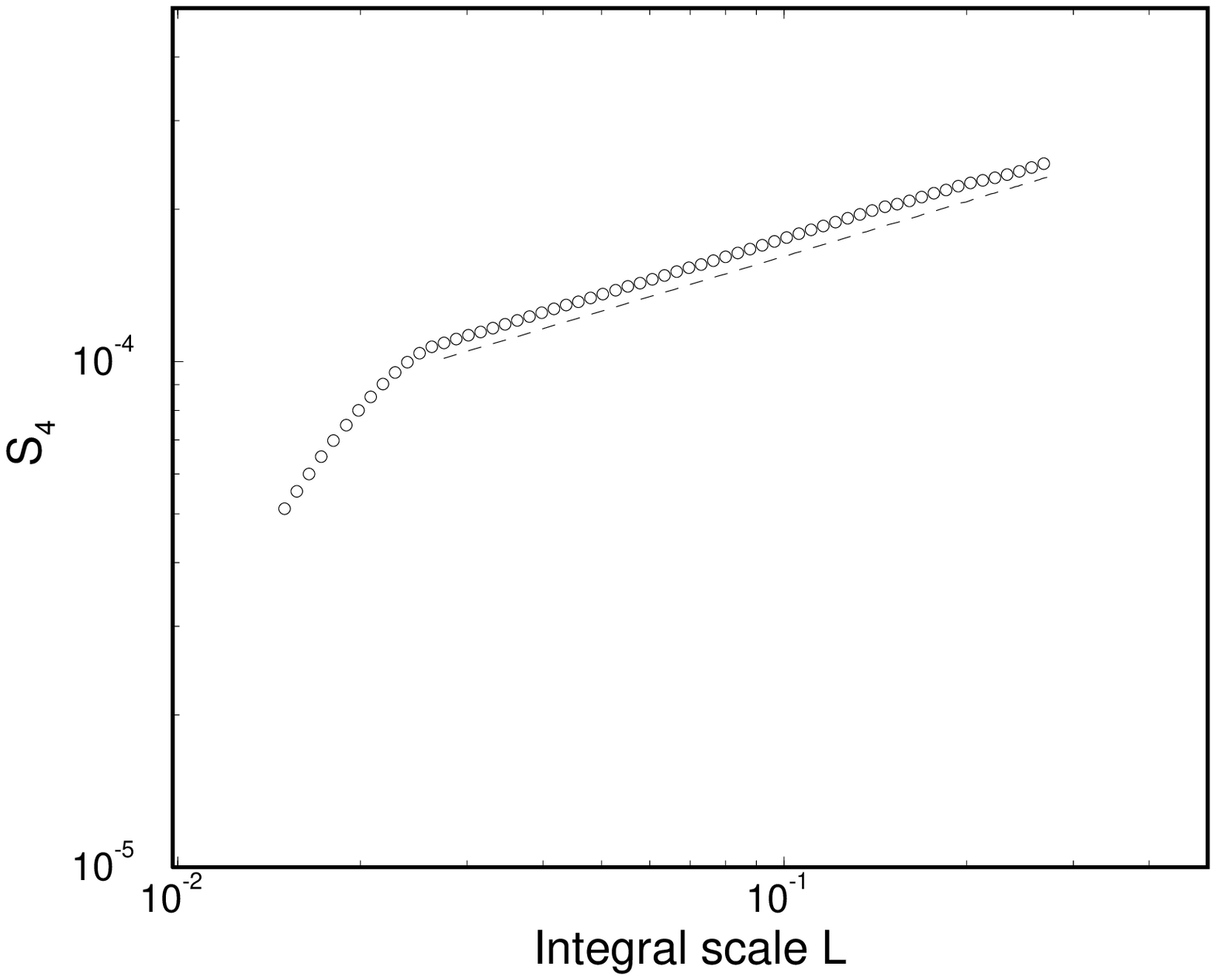,height=8cm,width=9cm}}
\end{center}
\vspace{-0.5cm}
\caption{Same curve as in Fig.~2 for $\xi=0.9$. The parameters are
$r=2.7\times 10^{-2}$, $\kappa=4.4\times 10^{-4}$ and the number of
realizations is $8\times 10^6$.}
\label{fig4}
\begin{center}
\mbox{\hspace{0.0cm}\psfig{file=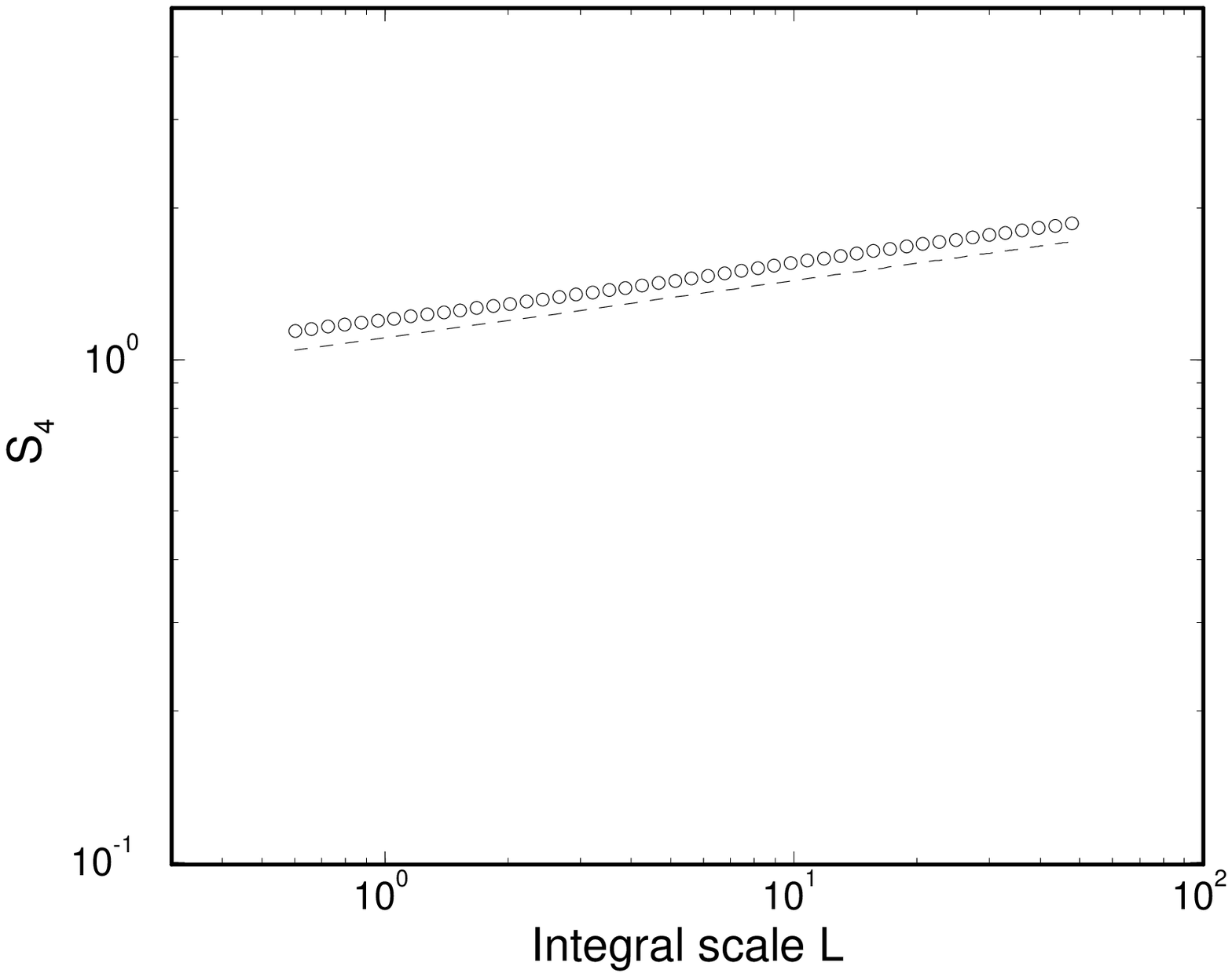,height=8cm,width=9cm}}
\end{center}
\vspace{-0.5cm}
\caption{Same curve as in Fig.~2 for $\xi=1.75$.
The parameters are $r=2.7\times 10^{-2}$,
$\kappa=10^{-9}$ and the number of realizations is
$1.5\times 10^6$.}

\label{fig5}
\end{figure}

\begin{figure}
\begin{center}
\mbox{\hspace{0.0cm}\psfig{file=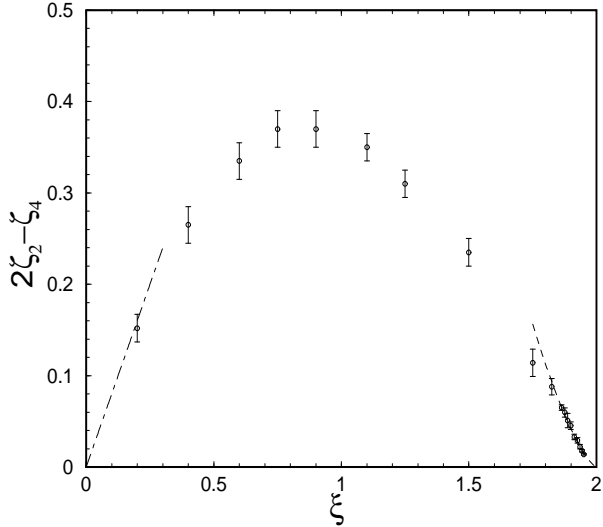,height=8cm,width=9cm}}
\end{center}
\vspace{-0.5cm}
\caption{The anomaly $2\zeta_2-\zeta_4$ for the
fourth--order structure function in the three-dimensional Kraichnan
model.}
\label{fig2}
\end{figure}

We applied our method to the determination of the anomalies
for the fourth-order structure function
$S_4(r\,;L)\propto r^{2\zeta_2}\left(L/r\right)^{2\zeta_2-\zeta_4}$ in
three dimensions.  The results are
summarized in the curve of the anomalous correction $2\zeta_2-\zeta_4$
{\it vs} $\xi$ presented in Fig.~\ref{fig2}.
The three plots of $S_4$ {\it vs} $L$ in Figs.~\ref{fig3}, \ref{fig4}
and \ref{fig5} indicate that the Lagrangian simulations require more
and more computational effort when $\xi$ is decreased from 2 to 0.
This is due mainly to the fact that the three correlation functions
appearing in the expression of $S_4$ have dominant contributions
scaling as $L^{2(2-\xi)}$ and $L^{2-\xi}$, but they are both cancelled
in the combination giving $S_4$. Making the subdominant contribution
of $S_4$ to emerge requires stronger and stronger cancellations as
$\xi$ decreases. For all the cases reported the scaling is quite
clean, as also confirmed by the analysis of local scaling exponents
(on octaves ratios), whose fluctuations give the conservative 
\noindent error bars in
Fig.~\ref{fig2}.

The dot-dashed line in Fig.~\ref{fig2} is the first-order
perturbative prediction $(4/5)\xi$, obtained in Ref.~\cite{GK95}. The
dashed line is a fit of the form $a \gamma + b \gamma^{3/2}$ with
$\gamma = 2-\xi$ (the parameters are $a=0.06$ and $b=1.13$), showing
that the data are compatible with an expansion in
$\sqrt{\gamma}$. Note that a term $\propto\sqrt{\gamma}$ is ruled out by
the H\"older inequality $\zeta_4 \leq 2\zeta_2=2\gamma$
\cite{gamma97}. It is interesting to remark that the crossing of the
curve in Fig.~\ref{fig2} with the monotonically decreasing (in $\xi$)
linear ansatz prediction occurs around $\xi=1$. This is the  point
farthest from the two limits $\xi=0$ and $\xi=2$ which both have strongly
nonlocal dynamics, suggesting a possible relation between deviations from
the linear ansatz and locality of the interactions \cite{RHK98}.

We finally note that  the two main ingredients of the method reported here
have in fact  a wider range of applicability than the determination of
anomalies for the Kraichnan model. First, the Lagrangian tracer method
appears more flexible than the integration of the passive scalar partial
differential equation.  The latter permits in principle measurement of all
the observables and somehow corresponds to an infinite number of tracer
particles. Changing their number according to which specific correlation
function is being investigated seems however to be more economic and
convenient and should  also be of interest for analyzing  the advection by
more realistic flow. Second, considering the scaling behavior {\it vs} the
integral scale $L$, rather than {\it vs} the separation $r$, could be
useful in many situations where the injection rate can be controlled;
this  includes the simulation of  Navier--Stokes flow with white-in-time
forcing. Such a procedure presents the  advantage of giving direct access
to the scaling exponent anomalies, which are quantitative measurements of
intermittency.

We are grateful to A.~Noullez for  useful
interactions throughout the course of this work. We acknowledge
valuable discussions with M.~Chertkov, G.~Falkovich, O.~Gat,
K.~Gaw\c{e}dzki, R.H.~Kraichnan, S.A.~Orszag, I.~Procaccia, A.~Wirth,
V.~Yakhot and R.~Zeitak.  Simulations
were performed in the framework of the SIVAM project of the
Observatoire de la C\^ote d'Azur. Part of this work was done while UF and
MV were visiting the Institut des Hautes \'Etudes Scientifiques.
Partial support from the  CNRS through a  ``Henri Poincar\'e'' fellowship
(AM) and from the GdR ``M\'ecanique des Fluides G\'eophysiques et
Astrophysiques'' (MV) is also acknowledged.


\begin{thebibliography}{99}
\bibitem{RHK94} R.H.~Kraichnan, {\it Phys. Rev. Lett.}, {\bf 52},
1016, (1994).
\bibitem{RHK68}
R.H.~Kraichnan, {\it Phys. Fluids}, {\bf 11}, 945, (1968).
\bibitem{UF95} U.~Frisch, {\it Turbulence. The Legacy of
A.N.~Kolmogorov}, Cambridge Univ. Press, Cambridge, (1995).
\bibitem{FGLP96}
A.L.~Fairhall, O.~Gat, V.S.~L'vov \& I.~Procaccia, {\it Phys.
Rev. E}, {\bf 53}, 3518, (1996).
\bibitem{RHK97}
R.H.~Kraichnan, {\it Phys. Rev. Lett.}, {\bf 78}, 4922, (1997).
\bibitem{CFKL95}
M.~Chertkov, G.~Falkovich, I.~Kolokolov \& V.~Lebedev, {\it Phys. Rev. E},
{\bf 52}, 4924 (1995).
\bibitem{GK95}
K.~Gaw\c{e}dzki \& A.~Kupiainen, {\it Phys. Rev. Lett.}, {\bf 75},
3834, (1995).
\bibitem{SS95}
B.I.~Shraiman \& E.D.~Siggia, {\it C.R. Acad. Sci.}, {\bf 321}, S\'erie II,
279, (1995).
\bibitem{nonlinear}Closures for the nonlinear Navier--Stokes equation can
also lead to anomalous scaling via ``fluxless'' solutions, such as
those found in dimension $d$ close to two
(U.~Frisch \& J.D.~Fournier {\it Phys. Rev. A}, {\bf 17}, 747 (1978)).
\bibitem{SS96}
B.I.~Shraiman \& E.D.~Siggia, {\it Phys. Rev. Lett.}, {\bf 77}, 2463, (1996).
\bibitem{KYC95}
R.H.~Kraichnan, V.~Yakhot \& S.~Chen,
{\it Phys. Rev. Lett.}, {\bf 75}, 240, (1995).
\bibitem{FGLP97}
A.L.~Fairhall, B.~Galanti, V.S.~L'vov \& I.~Procaccia,
{\it Phys. Rev. Lett.}, {\bf 79}, 4166, (1997).
\bibitem{GPZ97}
Using Monte--Carlo simulations of Lagrangian trajectories was
proposed independently by O.~Gat, I.~Procaccia \& R.~Zeitak, (1997),
unpublished\,; see also O.~Gat \& R.~Zeitak, ``Multiscaling in passive scalar
advection as stochastic shape dynamics'', cond-math/9711034.
\bibitem{BGK97}
D.~Bernard, K.~Gaw\c{e}dzki \& A.~Kupiainen, ``Slow modes in passive
advection'', cond-math/9706035. {\it J. Stat. Phys} (in press).
\bibitem{MC97}
M.~Chertkov, {\it Phys. Rev. E}, {\bf 55}, 2722, (1997).
\bibitem{KP92}
P.E.~Kloeden \& E.~Platen, {\it Numerical Solution of Stochastic Differential
Equations}, Springer (1992).
\bibitem{gamma97}
Similar H\"older inequalities rule out a $\sqrt{\gamma}$ term
for all the structure functions $\langle |\theta(r)-\theta(0)|^p\rangle$
with $p>2$. No such constraint exists for
$\langle\left(\theta(r)-\theta(0)\right)^3\rangle$, considered in
Refs.~\cite{BFL97,GLP97}; hence, the presence of an absolute value
strongly affects the scaling behavior.
\bibitem{BFL97}
E.~Balkovski, G.~Falkovich \& V.~Lebedev, {\it Phys. Rev. E}, R4881, (1997).
\bibitem{GLP97}
O.~Gat, V.~L'vov, E.~Podivilov \& I.~Procaccia, {\it Phys. Rev. E},
3863, (1997).
\bibitem{RHK98}
R.H.~Kraichnan, private communication, (1998).

\end{thebibliography}
\end{document}